\documentclass[prl,aps,twocolumn]{revtex4}
\usepackage{graphicx} 
\usepackage[usenames]{color}
\usepackage{amsmath,amssymb}
\usepackage{gensymb}
\usepackage{natbib}
\usepackage{xcolor}
\def\strutdepth{\dp\strutbox}
\def\nw#1{\strut\vadjust{\kern-\strutdepth\vtop to0pt{\vss\hbox to\hsize
{\hskip\hsize\hskip5pt$\leftarrow$\hss\strut}}}{\em #1}}
\usepackage{rotating}

\begin{document}

\title{A non-local rheology for granular flows across yield conditions}
\author{Mehdi Bouzid}
\author{Martin Trulsson}
\author{Philippe Claudin}
\author{Eric Cl\'ement}
\author{Bruno Andreotti}
\affiliation{Physique et M\'ecanique des Milieux H\'et\'erog\`enes, PMMH UMR 7636 ESPCI -- CNRS -- Univ.~Paris-Diderot -- Univ.~P.M.~Curie, 10 rue Vauquelin, 75005 Paris, France}

\begin{abstract}
The rheology of dense granular flows is studied numerically in a shear cell controlled at constant pressure and shear stress, confined between two granular shear flows. We show that a liquid state can be achieved even far below the yield stress, whose flow can be described with the same rheology as above the yield stress. A non-local constitutive relation is derived from dimensional analysis through a gradient expansion and calibrated using the spatial relaxation of velocity profiles observed under homogeneous stresses. Both for frictional and frictionless grains, the relaxation length is found to diverge as the inverse square-root of the distance to the yield point, on both sides of that point.
\end{abstract}


\date{\today}
\maketitle

Granular materials belong to the class of amorphous athermal systems. Like foams \cite{BW90,Kabla2003}, emulsions \cite{CCSB09}, suspensions \cite{BDCL10,BGP11,TAC12} or metallic glasses \cite{Xie08}, they exhibit a dynamical phase transition between \emph{static} and \emph{flowing} states. Analogously to phase transitions of thermodynamic systems,  this \emph{rigidity} transition exhibits a divergence of correlation lengths \cite{WNW05,HBB10}, revealing the presence of non-local cooperative processes called dynamical heterogeneities \cite{DJvH}. In order to describe the constitutive behavior of such systems, it is natural to adopt the Ginzburg-Landau phenomenological approach of phase transitions \cite{AT06,GCOAB08,BCA09,KK12,HK13}. The main issue is then to identify the relevant control and order parameters. Following the now classical Liu-Nagel diagram for jamming transition \cite{LL98} -- or a revised version \cite{BZCB11} -- it is usually assumed that the solid-liquid mechanical transition is controlled by  the shear stress $\tau$ \cite{GCOAB08,BCA09,KK12}, which, once rescaled by its critical value, defines the yield parameter ${\mathcal Y}$. For a granular system sheared under a fixed confining pressure $P$, one defines the dimensionless Coulomb yield parameter ${\mathcal Y}=\tau/(P \mu_c)$, where $\mu_c$ is the friction coefficient in the zero shear-rate limit, at the jamming volume fraction $\phi_c$.

\begin{figure}[t!]
\includegraphics{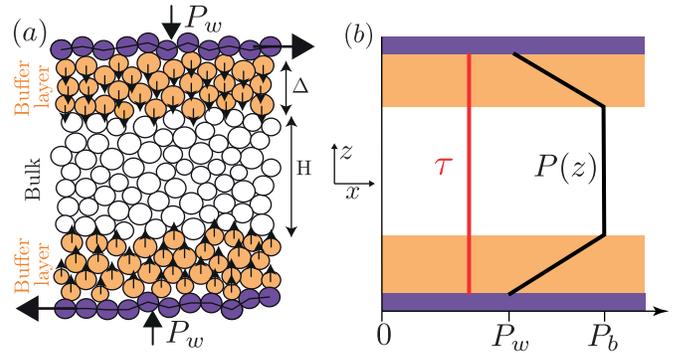}
\vspace{-4 mm}
\caption{
(Color online) (a) Schematic of the numerical set-up. The walls, composed by the dark purple grains, are submitted to a confining pressure $P_w$. In the buffer layers located close to the walls, the grains (orange) are submitted to gravity-like forces along the transverse direction $z$. (b) Schematic profiles of the pressure $P$ (black line) and of the shear stress $\tau$ (red line) across the cell. In the bulk of the shear cell (white), the pressure is homogeneous: $P=P_b$.}
\vspace{- 4 mm}
\label{Fig1}
\end{figure}

In a series of recent papers \cite{BCA09,GCOAB08,CMCB12,KK12,HK13} the order parameter is a rheological quantity called the fluidity, proportional to the inverse viscosity i.e. to the ratio of the shear rate $\dot \gamma$ to the shear stress $\tau$. Here, we consider that the relevant order parameter must be a dimensionless quantity based exclusively on state variables (which excludes $\tau$) like the shear rate $\dot \gamma$ rescaled by a microscopic timescale. In granular materials, the only energy scale is set by the confining pressure $P$, so that the order parameter must be the inertial number
\begin{equation}
I = \frac{|\dot \gamma| d}{\sqrt{P/\rho}},
\label{RheoLocal}
\end{equation}
based on the grain diameter $d$ and on their density $\rho$. $I$ compares $\dot\gamma$ to the microscopic rearrangement time $d\sqrt{\rho/P}$. Considering an incompressible homogeneous flow, it can be inferred that the yield parameter is a function of $I$ noted $\mathcal{Y}=\mu(I)/\mu_c$ \cite{GDRMidi,CEPRC05,JFP06}. If this local constitutive relation was still valid in heterogeneous flows,  the transition between solid ($I=0$) and liquid ($I>0$) states would systematically occur at ${\mathcal Y}=1$. However, different experiments have shown that the stress at a location depends on the shear rate \emph{around} this point, a property called \emph{non-locality}. (i) In the inclined plane geometry, thin granular layers flow anomalously \cite{DLDA06} and stop at a yield parameter ${\mathcal Y}>1$ \cite{P99,GDRMidi}. (ii) A creeping flow is commonly observed in regions which are expected to be jammed (i.e. solid), since ${\mathcal Y}<1$ \cite{KINN01,GDRMidi,BAU11}. (iii) A solid plunged in grains and submitted to a force lower than the yield threshold starts moving as soon as a shear band is created far away from the solid \cite{NZBWH10,RFP11}.

In this letter, we show that the liquid state continuously extends from liquid zones (${\mathcal Y}>1$) into the bulk of regions that are below the yield conditions (${\mathcal Y}<1$). We furthermore find that the rheology obeys the very same non-local constitutive relation across yield conditions.

Three different pictures have emerged so far to explain non-locality~\cite{A07}. In soft amorphous systems, like foams, emulsions or glassy Lennard-Jones phases, the dynamics in the quasi-static regime is controlled by elasto-plastic events~\cite{TLB06,LC09,BCA09,LP09}: when sheared, energy is slowly stored and rapidly released through scale-free avalanches, in close analogy with the depinning transition of an elastic line. By contrast, the dynamics of hard non-deformable grains is essentially related to geometry: elementary plastic events are rather identified as the rapid formation of force chains followed by a slow zig-zag instability of these structures~\cite{LDW12}. Non-locality can then be related to soft modes, by essence spread in space, prescribing the cooperative motion of the particles \cite{ABH12}. In this geometrical picture, the relevant state parameter would rather be the mean number of contacts per particle $Z$ or the volume fraction $\phi$ \cite{H10}. The third picture is based on an analogy with Eyring's transition state theory for the viscosity of liquids \cite{PF09}, where mechanical fluctuations would play the role of temperature in thermal systems. Here, we show that the non-local constitutive relation for dense granular flows can be determined from simple phenomenological assumptions, regardless the nature of the relevant dynamical mechanisms. We calibrate and test it by means of discrete element simulations.
\begin{figure}[t!]
\includegraphics{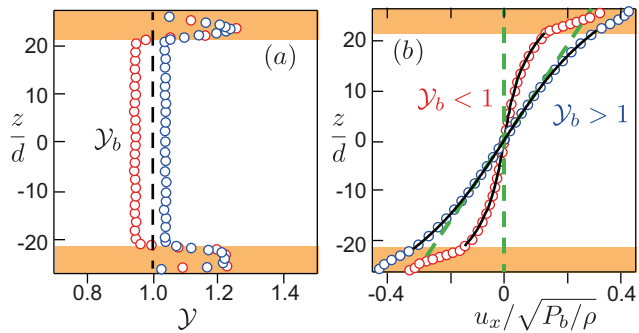}
\vspace{-6 mm}
\caption{(Color online) (a) Typical profiles of the yield parameter $\mathcal{Y}$ obtained numerically below (red circles, $\mathcal{Y}_b<1$) and above (blue circles, $\mathcal{Y}_b>1$) yield conditions for frictional grains.  (b) Corresponding velocity profiles. Green dashed lines: predictions of the local rheology. Black solid lines: best fits by Eq.~\ref{proloc}.}
\vspace{-4 mm}
\label{Fig2}
\end{figure}

\emph{Numerical set-up}~---~We have performed molecular dynamics simulations of massive grains, confined in a shear cell under an imposed stress field. The system is two-dimensional and constituted of $N \simeq 2\cdot 10^3$ circular particles of mean diameter $d$, with a $\pm$20\% polydispersity. The shear cell is composed of two rough walls moving along the $x$-direction with opposite velocities (see Fig.~\ref{Fig1}a for notations). These walls are made of  the same grains, but glued together. The walls are separated by $H+2\Delta \simeq 55 d$. Their position is controlled to ensure a constant normal stress $P_w$. The cell thickness then fluctuates over a fraction of grain diameter. Periodic boundary conditions are applied along the $x$-direction. The particle and wall dynamics are integrated using the Verlet algorithm. Contact forces between particles are modeled as linear viscoelastic forces, chosen such that the restitution coefficient is $e \simeq 0.9$. For each measurement reported here, we have varied  the normal spring constant $k_n$ and we report the value of the plateau at large $k_n$. This rigid asymptotic regime is reached in practice for $k_n/P>10^{3}$. To model actual granular samples, we have considered frictional grains, which interact along the tangential direction by a Coulomb friction \cite{Cundall79,daCruz05,Luding06} of coefficient $\mu_p=0.4$, with a tangential spring constant $k_t=0.5 k_n$. For the sake of comparison, we have also studied the same system with frictionless grains ($\mu_p=0$).

What makes the set-up original is the possibility of imposing the profile of $\mathcal{Y}(z)$ by means of gravity-like forces applied to the grains located in two buffer zones (Fig.~\ref{Fig1}a) of thickness $\Delta=5d$. A grain labelled $i$, of mass $m_i$, and located at $z=z_i$ is submitted to an external force:
$f_z^i =  m_i g \left[
- e^{ -\frac{(z_i - (H+ \Delta)/2)^2}{2d^2} }
+ e^{ -\frac{(z_i+(H + \Delta)/2)^2}{2d^2} }
\right]$,
where $g$ is the amplitude of the localized "gravity" field. These forces are oriented downward at the top of the cell, and upward at the bottom (Fig.~\ref{Fig1}a). The resulting pressure $P$ (Fig.~\ref{Fig1}b)  starts from $P_w$ at the upper wall, increases due to the sum of forces applied in the buffer zone, and reaches a constant value $P_b$ in a central region of width $H$, called the bulk. $P$ decreases back to $P_w$ at the lower boundary. By contrast, the shear stress profile is homogeneous across the cell (Fig.~\ref{Fig1}b). As a result,  the stresses are homogeneous  in the bulk of the shear cell: $\mathcal{Y}=\mathcal{Y}_b$. By tuning the amplitude $g$,  $\mathcal{Y}_b$ can be imposed smaller or larger than $1$. $P_w$ is chosen larger than $\tau/\mu_c $ so that the buffer zone remains above yield conditions, at $\mathcal{Y}>1$.

\emph{Local rheology}~---~For given stress conditions, the simulation is ran until a steady state is reached and the averaged velocity profiles are then measured. Fig.~\ref{Fig2} shows two such profiles, one above and the other below yield conditions. One observes that the entire system always flows, even when $\mathcal{Y}_b<1$. In the bulk, the velocity deviates from the linear profile $u_x=\dot \gamma_\infty z$ predicted by the local rheology (as $\mathcal{Y}$ is constant, one expects $I$ to be constant as well and to vanish for $\mathcal{Y}_b<1$). The velocity rather tends exponentially towards such a linear profile, which suggests a linear relaxation in space. The velocity profiles inside the bulk zone (Fig.~\ref{Fig2}) are accordingly fitted with the function:
\begin{equation}
u_x(z)=\dot\gamma_\infty z + \frac{u_x(H/2)-\dot\gamma_\infty H/2}{\sinh(H/(2\ell))}\;\sinh(z/\ell)\label{proloc}
\end{equation}
The velocity $u_x(H/2)$ is inherited from the buffer layer, while the asymptotic shear rate $\dot\gamma_\infty$ (which is \emph{not} the shear rate at the center of the cell) and the relaxation length $\ell$ are two adjustable parameters. Varying $\mathcal{Y}$ in the buffer layer but keeping $\mathcal{Y}_b$ constant, we systematically measured the same values of $\dot\gamma_\infty$ and $\ell$. These two quantities characterize the bulk state and do not depend on the buffer layer characteristics.

\begin{figure}[t!]
\includegraphics{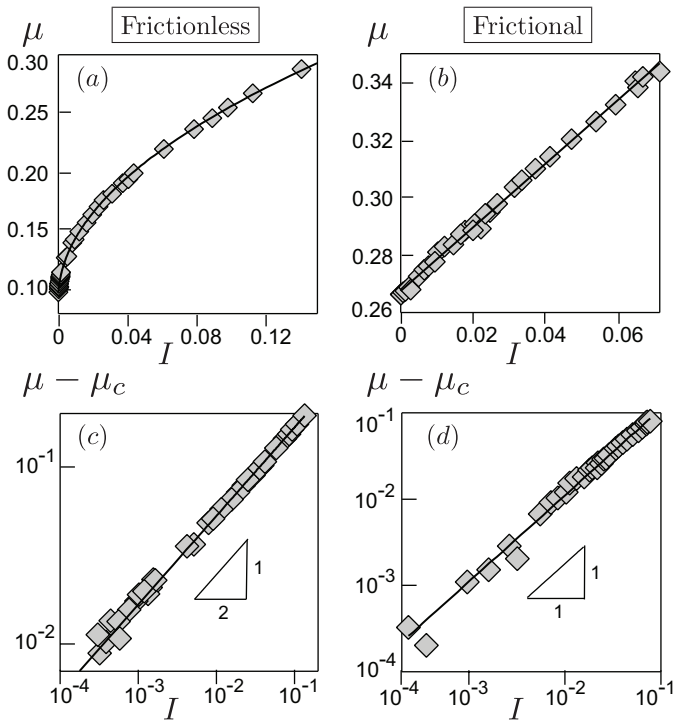}
\vspace{-4 mm}
\caption{Local rheology $\tau/P=\mu(I)$ deduced from the fit of the bulk velocity profile to Eq.~\ref{proloc}. Data for frictionless (a) and frictional (b) grains. Solid lines: the best fit to Eq.~(\ref{mupow}) gives $\mu_c=0.094, \alpha=0.5, b=0.518$ for the frictionless case and $\mu_c=0.267, \alpha=1.0, b=1.148$ for the frictional one. Bottom: $\tau/P - \mu_c$ as a function of $I$ in log scales for the frictionless (c) and frictional (d) cases.}
\vspace{-4 mm}
\label{Fig3}
\end{figure}

The asymptotic shear rate $\dot\gamma_\infty$ provides the proper way of defining the local rheology $\mu(I)$.  $I = |\dot \gamma_\infty| d/\sqrt{P/\rho}$ is indeed the inertial number selected for a certain ratio $\tau/P$ in homogeneous shear {\rm and} stress conditions. It is deduced in practice from the fit of the data to Eq.~\ref{proloc}. The resulting constitutive relations are reported in Fig.~\ref{Fig3} for the frictional and the frictionless case. In both cases, the data is perfectly described by a law of the form
\begin{equation}
\mu = \mu_c + b I^\alpha\label{mupow}
\end{equation}
in the accessible range of $I$ (between $10^{-4}$ and $10^{-1}$): the residuals form a statistical noise. The exponent $\alpha$ is $0.5$ in the frictionless case and $1$ in the frictional case, within error bars ($\sim 5\%$), see also \cite{CEPRC05,PR08}. We hypothesize that this difference is related to another fundamental difference between the two systems. In the frictionless situation, the jamming point ($I=0$ and $\phi=\phi_c$) coincides with the isostatic point, while for frictional grains, the jamming point is far in the hyper static zone \cite{WNW05,H10}.
%

\begin{figure*}[t!]
\includegraphics{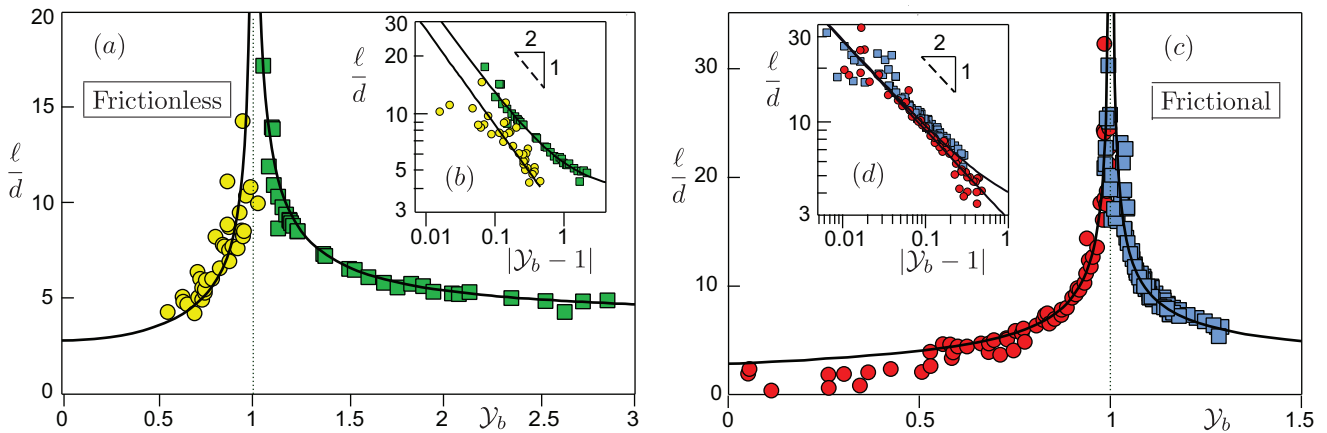}
\vspace{- 4 mm}
\caption{
(Color online) (a) Relaxation length $\ell$ as a function of $\mathcal{Y}_b$, below (circles) and above (squares) yield conditions. (a) Green and yellow symbols: data for frictionless grains. (c) Blue and red symbols: data for frictional grains. The solid lines are the best fit by Eqs.~(\ref{ellabove}) and (\ref{ellbelow}). The values of $\nu$ found for frictionless ($\nu=7.83 \pm 0.21$) and frictional ($\nu=8.08 \pm 0.49$) systems are remarkably similar. (b) and (d) Log-log plots of the same quantities, revealing the divergence with an exponent $-1/2$.}
\vspace{-4 mm}
\label{Fig4}
\end{figure*}

\emph{Non-local rheology}~---The relaxation length $\ell$ is displayed in Fig.~\ref{Fig4} as a function of  $\mathcal{Y}_b$. The numerical data for both frictionless and frictional systems are qualitatively similar: $\ell$ diverges on both sides of the critical point $\mathcal{Y}_b=1$ with an exponent $1/2$. In order to account for non-local effects in the theoretical constitutive relation, we perform a gradient expansion of the functional $\mathcal{Y}[I]$. Assuming that non-locality results from a statistically isotropic short-range interaction between shear zones, the lowest order operator is the Laplacian $\nabla^2 I$. As a direct consequence, $I$ and its gradient must be continuous. Furthermore, we assume that the correction remains finite as $I\to 0$, so that the expansion must be expressed in terms of $\kappa \equiv d^2{(\nabla^2 I)}/{I}$. At the linear order in $\kappa$, the constitutive relation writes:
\begin{eqnarray}
\mathcal{Y}=\frac{\mu(I)}{\mu_c} \left[1-\nu\kappa\right]\, ,
\label{RheoNonLocal}
\end{eqnarray}
where $\nu$ is a phenomenological constant. $\kappa$ is positive when the point considered is surrounded by a more liquid region (higher $I$). This region flows more easily than expected from the local value of $I$, so that the corresponding shear stress is lower. $\nu$ is therefore positive. Importantly, our derivation does not depend on the nature of the mechanical interaction between shear zones; the reader may think of the analogy with the van der Waals gradient expansion of the Helmholtz free energy at a liquid-vapour interface \cite{Rowlinson}.

Above yielding conditions, the linearization of Eq.~\ref{RheoNonLocal} around the bulk inertial number $I = I_b + \delta I$ gives at first order a differential equation of the form $\ell^2 \frac{d^2 \delta I}{dz^2}-\delta I=0$, whose solutions are exponentials with a relaxation length
\begin{equation}
\ell_>=d\;\sqrt{ \frac{\mathcal{Y}_b \nu}{\alpha(\mathcal{Y}_b-1)}} \quad{\rm for}\quad \mathcal{Y}_b>1.
\label{ellabove}
\end{equation}
Below yielding conditions, as $I_b=0$, the non-local correction is of zeroth order and Eq.~\ref{RheoNonLocal} leads to $\kappa =(1-\mathcal{Y}_b)/\nu$. This gives a similar differential equation but now with a divergence of the form
\begin{equation}
\ell_<=d\;\sqrt{\frac{ \nu}{1-\mathcal{Y}_b}} \quad{\rm for}\quad \mathcal{Y}_b<1.
\label{ellbelow}
\end{equation}
As shown in Fig.~\ref{Fig4}, the measured relaxation length $\ell$ effectively diverges on both sides of the critical point according to the theoretical predictions (\ref{ellabove}, \ref{ellbelow}). In particular, both frictional and frictionless systems exhibit a power-law divergence as $\sim |\mathcal{Y}_b-1|^{-1/2}$; the multiplicative factor is the same above and below $\mathcal{Y}_b=1$ for the frictional case but differ by $\sqrt{\alpha}$ in the frictionless case (Fig.~\ref{Fig4}b and \ref{Fig4}d). Note that a similar scaling was found in \cite{BCA09}  for a fluidity correlation length in a kinetic elastoplastic model.

\emph{Discussion}~--- 
It must be emphasized that the creeping regime $\mathcal{Y}<1$ and in the fully flowing regime $\mathcal{Y}>1$) are described as a {\it single liquid phase}. The divergence of the relaxation length is indeed predicted across the yield condition by the very same non-local correction to the liquid constitutive relation. Due to the liquid boundary condition imposed by the buffer layers, the bulk is always found flowing, regardless of the stress in this zone. The liquid state persists asymptotically into the bulk, far from any direct influence of the boundary. This is consistent with our choice of order parameter: $I$ is everywhere non zero and the same constitutive relation holds in all layers. In conclusion, the liquid-solid transition is not controlled by $\mathcal{Y}$. For $\mathcal{Y}$ smaller than static threshold value $\mathcal{Y}_s$ the system can be solid or fluid. If fluid then $I$ varies in space according to the non-local rheology. If solid, the stress state must be described by another constitutive relation based on elasticity.

Our derivation, based on a gradient expansion of the yield parameter $\mathcal{Y}[I]$, written as a functional of the order parameter $I$, does not  prejudge of any dynamical mechanisms at work at the microscopic level. The exponential behaviors and the associated length-scales hence identified are direct consequences of the linearization around the critical state. Therefore, finer investigations at the grain level must be carried out to understand the connections between the three lines of thought currently invoked to explain such a non-local rheological coupling, namely elasto-plastic \cite{TLB06,LC09,BCA09,LP09}, geometrical  \cite{WNW05,LDW12} and stress-mediated activation \cite{PF09,RFP11}. 

BA is supported by Institut Universitaire de France. This work is funded by the ANR JamVibe.

\vspace{-4 mm}

\end{document}